**Microsphere solid-state biolasers**

*Van Duong Ta[+], Soraya Caixeiro, Francisco M. Fernandes and Riccardo Sapienza\**

Dr. V. D. Ta, S. Caixeiro, Dr. R. Sapienza
Department of Physics, King's College London, Strand, London WC2R 2LS, UK
+ E-mail: van_duong.ta@kcl.ac.uk
\*E-mail: Riccardo.Sapienza@kcl.ac.uk

Dr. F. M. Fernandes
Laboratoire de Chimie Matiere Condensée de Paris, Université Pierre et Marie Curie, Sorbonne Universités, UMR7574, 4 Place Jussieu, 75005 Paris, France

Optical biosensors, which are label-free and sensitive, have been widely used in the life sciences.[1] In the last decade, the passive whispering gallery mode (WGM) resonators have significantly improved the sensor sensitivity, reaching down to the single molecule level.[2-4] In these WGM biosensors, a tapered fiber is an important component that serves to drive the resonance mode and to guide the analyte induced wavelength shift. However, the presence of the fiber makes the sensing system complex. Alternatively, active WGM devices (lasers) based sensors are more convenient where the tapered fiber is not required. The laser is obtained by free-space excitation and the lasing emission is directly used for sensing applications such as refractive index change in the medium outside a cavity.[5, 6] WGM lasers have been recently integrated to living cells for intracellular sensing and cell tracking, which opens a new frontier of biotechnology.[7, 8]

WGM lasers made of polymers of biological origin are of interest for biosensing. To date, various biolasers have been investigated using random cavity,[9-13] distributed feedback (DFB),[14-16] Fabry-Perot (F-P)[17-19] and WGM capillary tube.[20] Generally, current devices are large (>>100 µm) and mainly rely on a biocompatible gain and a traditional solid-state cavity. A great effort has been spent on minimizing the laser size and developing fabrication techniques that use all-biocompatible materials. This has been achieved in fluid droplet lasers with cavity size smaller than 40 µm.[21] Instead, solid-state all-biomaterial lasers are limited to millimeter-scale size, which is too large for practical integration with living tissues/cells.[22] Unlocking the full potential of biointegration requires miniature lasers made of biocompatible and biodegradable materials for future implantable biosensing.[23, 24]

In this work we demonstrate miniature (15-100 µm) solid-state biolasers based on biomaterial cavities. We report on two classes of biopolymers, proteins such as Bovine Serum Albumin (BSA), and polysaccharides, such as pectin and cellulose.

Albumins are a family of globular proteins responsible for the transport of biomolecular species such as lipids, hormones, metal cations and therapeutic drugs in the blood stream and for the oncotic pressure maintenance in blood plasma.[25] In addition to their natural occurrence in mammals, albumins present high solubility in aqueous media and optical transparency, making them ideal components for the elaboration of optical devices.[26] Pectin and cellulose are biodegradable and sustainable biosourced polysaccharides derived from land plants.[27] Besides their key roles in plant growth, morphology and development,[28] these polysaccharides have recently picked up considerable attention for the elaboration of a wide range of biomaterials and biocompatible devices.[29-32] With regards to WGM cavities, the simplest structure is probably a dielectric sphere. It has intrinsically high quality (Q) factor and small mode volume,[33] which is appropriate for low threshold microlasers and sensors.[34-36] Microspheres lasers made of biologically produced materials are of interest for implantable biosensing as they are expected to offer excellent biocompatibility in medical applications. [11]

    We used BSA, pectin and cellulose (structure and chemical formula are shown in Figure S1) as material constituent of a spherical dielectric cavity capable of sustaining lasing. The fabrication of solid biopolymer microspheres is schematically shown in **Figure 1**. Firstly, biopolymer droplets were dispersed in an uncured PDMS resin (Base of Sylgard 184 Silicon Elastomer, Dow Corning) by emulsion processing similar to our previous work.[37] As water and PDMS are immiscible, aqueous droplets (Figure S2) can be spontaneously formed by mechanically disrupting the aqueous biopolymer-rich domains using a sharp needle (Figure 1a and b). Secondly, the droplets were subsequently annealed at 100 °C for dehydration and subsequent solidification (Figure 1c). The annealing time was 10, 60, 90 min. for BSA, pectin, cellulose spheres, respectively. Finally, the obtained microspheres were separated from PDMS by mixing in ethyl acetate-a relatively safe solvent (Figure 1d). As PDMS base resin is well dissolved in ethyl acetate the solid spheres can be easily separated from the liquid phase. Since ethyl acetate has a relatively high vapour pressure at ambient temperature, the evaporation of the remaining solvent was left to proceed at room temperature for several hours before using for characterizations. It is important to note that the PDMS does not only help to produce liquid droplets but also allows the formation of solid spheres. Generally, water droplets can be made using common oils but in that case it is very hard to evaporate the water and solidify the droplets to obtain solid spheres.

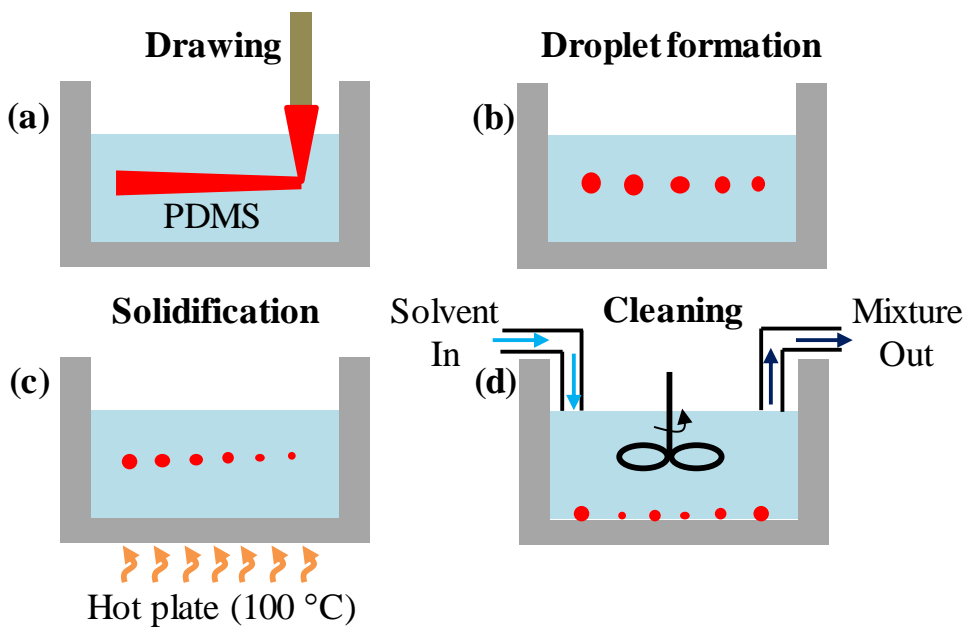

**Figure 1.** a)-d) Schematic diagram shows fabrication process of biomaterial microspheres.

The fabrication technique is flexible and a variety of cavity size can be produced by changing the droplet volume and density. **Figures 2**a and b show optical and SEM images of dye doped BSA microspheres with different sizes. Figure 2c plots the size distribution of 80 BSA spheres. The broad cavity size dispersion from 15 to 120 µm is a result of the manual fabrication process which could be improved by an automatic deposition, as for example by ink-jet printing.[38] The size-distribution presented can be controlled during the emulsion processing by changing droplet volume.

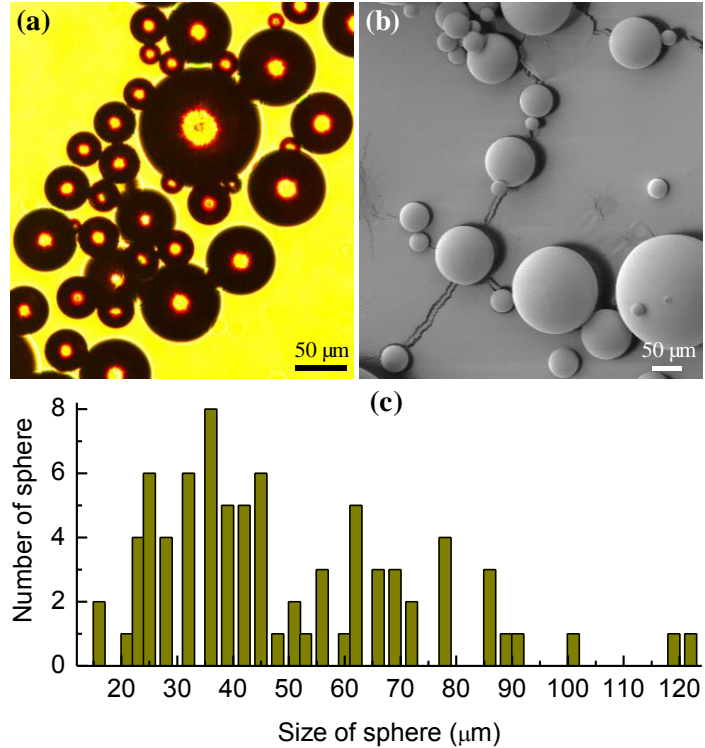

**Figure 2.** a) and b) Optical and SEM images of dye doped BSA microresonators, respectively. c) Size-distribution histogram of 80 BSA microspheres.

WGM lasers rely on multiple total internal reflection (TIR) at the interface between the cavity and the surrounding medium. When the TIR occurs, the light is trapped inside the cavity by resonant recirculation. After traveling one circumference if the light reaches the original point with the same phase then the resonance condition is fulfilled and a light mode is constructed, which is referred to as WGMs.[39]

We observe experimentally WGM lasing from the fabricated dye doped microspheres, upon pumping with a high-energy laser, as shown in **Figure 3**. The cavity loop is expected to form in the equator of the sphere (Figure 3a) and the emission can couple out the sphere by scattering which is subsequently collected by the microscope and measured by the spectrometer. The gain medium we use is Fluorescein, a well-known biocompatible dye, and Rhodamine B (RhB) which have been incorporated into the protein. The majority of the measurements reported here are performed in air.

The fluorescence-to-lasing transition is best observed in the power-dependent spectra as shown in Figure 3b. Under low pumping density (< 27 µJ/mm$^2$), the spectrum exhibits a spontaneous emission characterized by the weak intensity and broad linewidth. When the excitation density reaches 27 µJ/mm$^2$, sharp peaks with wavelength from 560 to 575 nm start to appear. As the pumping density increases, the output energy increases rapidly. The integrated fluorescence intensity shows a nonlinear increase versus excitation density (Figure 3c), indicating a distinct lasing threshold, which is about 20 µJ/mm$^2$. This threshold value for Fluorescein is higher than that of RhB doped spheres (1 µJ/mm$^2$) because the excitation wavelength (532 nm) is far from the optimized value of ~495 nm (the maximum absorption of Fluorescein). Indeed, when using RhB for the gain material, the lasers have a much lower threshold (~1 µJ/mm$^2$) as they can be more efficiently pumped.

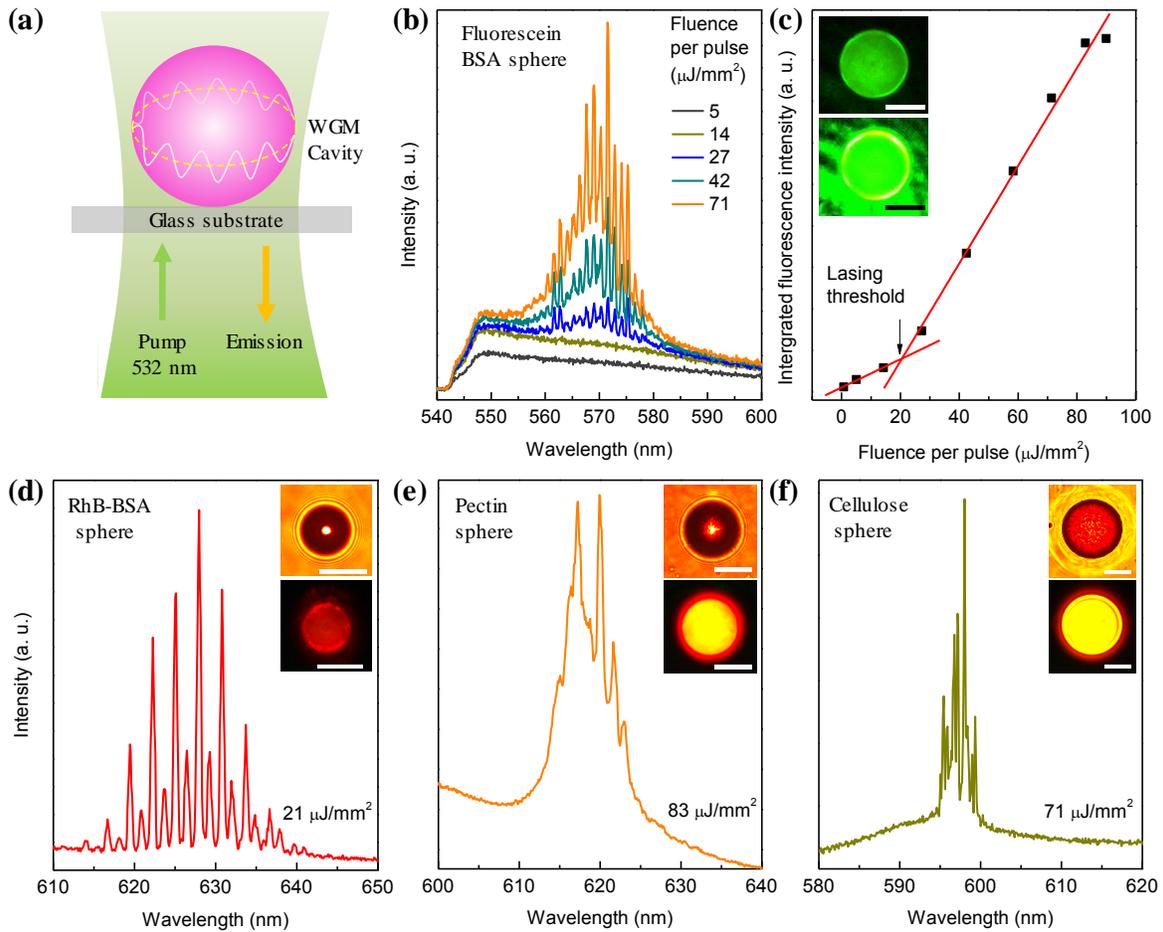

**Figure 3.** a) Schematic illustration of the optical measurement using a microscopy. The pumping and detecting light are performed through the same optical pathway using either 10 or 20x objective. Under optical excitation, whispering gallery mode (yellow line) is formed in the equator of the sphere. b) The lasing spectra obtained from a fluorescein doped BSA sphere as function of pumping energy density. c) The corresponding integrated fluorescence intensity versus excited fluence. The inset presents fluorescence images of the studied sphere that is below and above lasing threshold. The bright ring observed above the threshold shows the evidence of WGMs. d-f) Lasing spectra from typical RhB doped BSA, pectin and cellulose spheres, respectively. The insets show optical and fluorescence image of corresponding spheres. All scale bars are 30 µm.

SEM inspection indicates that the obtained microspheres have round shapes and diameters in the range of 15-100 µm (Figure S3). These biomaterial spheres lase with strong output emission upon optical pumping. Lasing spectrum of a typical RhB doped BSA, pectin and cellulose sphere, is plotted in Figures 3d-f where the lasing modes are well recognized above the fluorescence background. With regards to the lasing threshold, the cellulose sphere has the highest value of 24 µJ/mm$^2$ while the BSA sphere has the lowest of 1 µJ/mm$^2$ (Figure S4). The high threshold obtained from the cellulose sphere is attributed to a high optical loss due to scattering of the sphere which appear to be rougher at SEM image (Figure S3).

Generally, the lasing threshold increases with the decreasing sphere's size. As a result, the smallest lasing spheres would depend on the maximal allowed pumping energy. In our measurement, with pumping energy up to ~100 µJ/mm$^2$, the smallest spheres that lase are about 20 µm in diameter for BSA and 40 µm for pectin and cellulose.

The BSA spheres look very smooth in the SEM images and exhibit the clearest mode spacing which is convenient for investigation of the lasing mechanism. **Figure 4**a presents a typical lasing spectrum from the ~32 µm-diameter RhB-BSA sphere (the image is shown in the inset of Figure 3d). The spectrum exhibits two separated envelopes with different intensity. The lasing envelope and mode position can be well-explained by WGM theory. It is well-known that WGMs can be characterized by radial mode (*r*), angular or azimuthal mode (*m*) and its polarization, either transverse electric (TE) or transverse magnetic (TM) modes (schematic of TE and TM modes is shown in Figure S5).[40] For a droplet cavity, TE modes are generally expected to have higher Q factor than TM modes.[41] With this assumption, the lasing envelope with higher intensity is referred to TE modes while the other is TM modes. Consequently, lasing frequencies can be well-fitted by the asymptotic solutions of Mie scattering formalism.[40] The TE modes (from the left to the right) are well matched with mode numbers of 211 to 218 and TM modes are indexed as 211-216 (all modes are first radial modes *r* = 1 as they have the highest Q value) with assuming diameter of the sphere is *D* = 31.53 µm, refractive index of the sphere ($n_1$) and surrounding medium ($n_2$) are 1.42 and 1, respectively. The spectral linewidth (Δ*λ*) of the lasing modes is as small as 0.2 nm (which is limited by the resolution of our spectrometer), corresponding to a Q factor, defined as Q = *λ*/Δ*λ*, of approximately 3×10$^3$. This Q value is ~3 times higher than that of a solid-state protein ring lasers recently reported.[22]

In order to better verify the WGM mechanism, the free spectral range (FSR) of microlasers with different size was investigated. As shown in the inset of Figure 4b, the FSR of 23 and 38 µm-diameter spheres are 3.9 and 2.4 nm. Considering the resonant wavelength *λ* is about 625 nm then calculated FSR = *λ*/π$n_1$*D* for the two above spheres are 3.8 and 2.3 nm, respectively. The result shows an agreement between the theory and experimental observation, further verifying the WGM mechanism. Furthermore, as *λ* is similar for different spheres, the FSR of spheres of different size should follow a α/*D* function, where *α* is a constant. This is clearly seen in Figure 4b where the FSR as a function of sphere diameter is well-fitted by the function 89.2/*D* (*α* = 89.2).

To get a better understanding of WGMs, we have simulated the optical mode of the ~32 µm-diameter BSA sphere using the finite element method (FEM) supported by COMSOL Multiphysics.[42] The simulated results are presented in Figure S6 where the optical confinement of WGM can be clearly seen.

The lasing operation of BSA spheres is highly stable in ambient conditions. As shown in Figure S7, a BSA sphere continues lasing for as many as ~3×10$^4$ excitation pulses. The sphere also lases in an aqueous medium. Figure S8 shows that the lasing mode and intensity of an individual sphere maintain nearly the same values when it was immersed in water for 22 hours. Given our limited pump the minimum size of lasing BSA spheres in water is about ~30 µm. Furthermore, BSA spheres kept in culture conditions i.e. DMEM at 37 C with 5% $CO_2$ for more than a month continued to lase with a minor degradation of the threshold. In contrast, cellulose and pectin spheres tend to hydrate, swell and dissolve in water after several hours (cellulose) or a week (pectin). Consequently, we did not observe lasing from those spheres when they were in water.

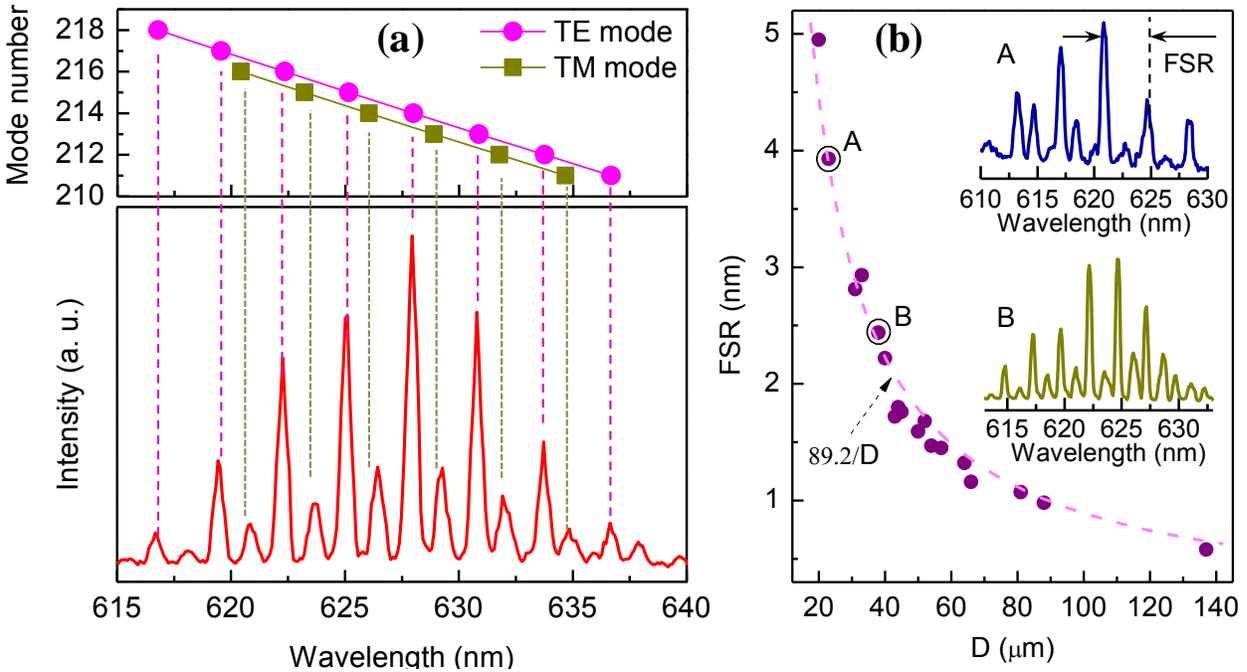

**Figure 4.** a) Matching between experimental observation of lasing wavelengths and calculated lasing modes. b) Free spectral space (FSR) as a function of sphere diameters. The insets show spectra of two typical spheres with diameter of ~23 (A) and 38 µm (B).

Biocompatibility can be considered as one of the most important factor in implantation in living tissues. To qualitatively assess the biocompatibility of our lasers, a simple experiment was carried out based on the replication rate of the HeLa when incubated in presence of biolaser microparticles. At least 50 spheres per culture well were immersed in a culture medium with HeLa cells and incubated for up to 4 days. **Figure 5** shows a BSA sphere microlaser surrounded by a number of HeLa cells after 3 days. The presence of BSA spheres did not prevent cell growth and division and no inhibition area was observed surrounding the laser particles. When compared to a control sample without spheres, the cells were found to split at similar rates and reached confluence after 4 days. While this is an indication of biocompatibility, a full cytotoxicity study should be conducted once a specific cell type of interest is identified for a specific biomedical application. In addition, during the 4 days of measurement, BSA spheres continued to lase in the culture medium with only a minor decrease in the lasing intensity.

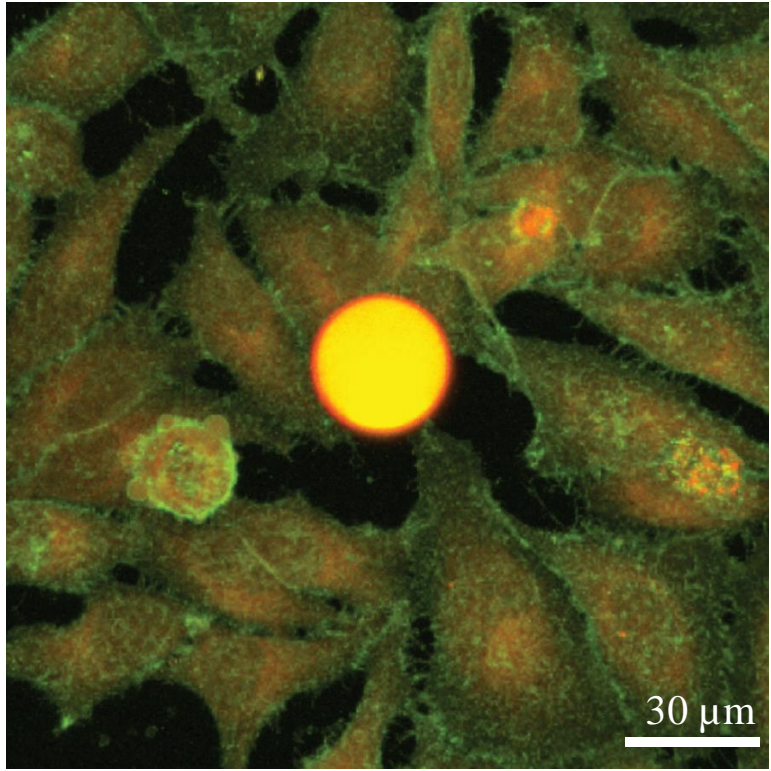

**Figure 5.** Confocal 3D scan (top view) of a biocompatible microlaser made of a RhB doped BSA microsphere (in orange) surrounded by Hela cells after 3 days of incubation. The cells were fixed with 5% Formaldehyde and the cellular membrane was stained with Alexa Fluor 488.

In conclusion, we have demonstrated solid-state miniature biolasers based on natural proteins and polysaccharides. Laser size ranges from 15 to 100 µm. The lasing mechanism is attributed to WGMs. Owing to low optical loss, the lasing threshold is only a few µJ/mm$^2$, which is as low as reported intracellular droplet lasers based on polystyrene cavities.[8] With Q factors of lasing modes as high as ~3×10$^3$, these lasers have potential for future ultrahigh-sensitive sensors. BSA microlasers were found to be resistant to the aqueous environment of cell culture media and biocompatible to cell growth. Stability in biological condition could be improved using a specific coating, as for example in the form of a silica layer. The fabrication process relies on aqueous solutions thus it is flexible and compatible with other biomaterials and other biocompatible gain media like vitamins [19] or Green Fluorescent Protein (GFP), [16] enabling different kinds of all-biomaterial solid-state microlasers. Due to the simplicity of operation and ease of fabrication our biolasers are promising for future biosensing applications.

## Experimental Section

**Biopolymer solution preparation**: Bovine serum albumin (BSA, lyophilized powder, ≥96%) was purchased from Sigma-Aldrich while pectin with high acetylation degree (extracted from beetroot) and cellulose (HPC, $m_w \sim 10^6$) were provided by our collaborators in France and Cambridge, UK, respectively. BSA was dissolved in deionized water with concentration of 1g/1mL by leaving the mixture in a fridge (~4 °C) for about 24 hours.[43] Pectin (4 wt%) and cellulose (1 wt%) were obtained by magnetic stirring the materials in water at room temperature overnight. Gain material (Fluorescein and Rhodamine B, Sigma-Aldrich) were then added by mixing 1 wt% aqueous laser dyes with the solvated biopolymers. The dye concentration in final mixtures (compared with the mass of the biopolymer) is approximately 0.8% (Fluorescein-BSA), 0.33% (RhB-BSA), 1% (RhB-Pectin) and 2% (RhB-Cellulose).

**Optical and SEM characterizations**: Optical measurement of individual microspheres was performed using a modified microscope system. The pumping source was a pulsed microchip Nd:YAG laser (from Teem photonics) with wavelength of 532 nm and pulse width of 400 ps. Pulse energy was controlled by an acousto-optic modulator (AOM). For laser experiment, the frequency used was up to 1 Hz. The sample was excited at normal incidence with beam spot of 165 µm using an objective lens with magnification of 10X (numerical aperture of 0.25 and working distance of 10.5 mm). The emission from the sample was collected through the same objective and was delivered to a spectrometer for spectra recording. The spectral resolution is ~0.2 nm. The spectra were integrated over a single pulse only. All experiments were carried at room temperature and in ambient conditions. Furthermore, surface morphology of microspheres was studied by scanning electron microscopy (SEM, Tescan). The spheres were coated with a thin gold layer by mean of thermal evaporation prior SEM analysis.

**Cell culturing**: HeLa cells were cultured routinely in vented flasks at 37 °C with 5% $CO_2$ in full growth medium without phenol red (Dulbecco's modified Eagle medium, DMEM, supplemented with fetal bovine serum 10%v/v and penicillinstreptomycin 1%v/v). Cells were passed when the culture reached ~85% confluency. Cell were detached from the flask with room temperature Trypsin (10 min, 37°C), washed with DMEM and 10% was suspended in a new flask with DMEM. BSA microsphere *in vitro* biocompatibility was assessed by incubating the particles with HeLa cells seeded in a sterile chambered coverslip, µ-Slide 8 well (Ibidi Labware) tissue culture treated under the previously described culture conditions. All optical characterization of cell-containing samples were conducted under ambient conditions for no longer than 1h after removal from the incubator to minimize cell damage.

**Cell imaging**: Cells, previously incubated with BSA spheres, were fixed and stained immediately prior to imaging. They were incubated with 5% Formaldehyde for 10 min, washed several times with PBS and subsequently stained with Alexa Fluor 488 for 10 min, and washed repeatedly with PBS. Images of each channel were captured with a Nikon Eclipse Ti-E Inverted microscope using a 40X Plan Apo objective. The green channel represents 488 nm excitation and 525nm emission, while the red channel represents 562 nm excitation and 595nm emission. Images were processed with NIS-Elements.

## Acknowledgements

This work was funded by the Engineering and Physical Sciences Research Council (EPSRC) under grant EP/M027961/1, the Leverhulme Trust (RPG-2014-238), and the Royal Society (RG140457). The data is publicly available in Figshare [44], the code is available on request.